\documentclass[%
 reprint,
 amsmath,amssymb,
 aps,
]{revtex4-2}

\usepackage{graphicx}
\usepackage{dcolumn}
\usepackage{bm}

\newcommand{\fig}[1]{Fig.~\ref{#1}}
\newcommand{\eq}[1]{Eq.~(\ref{#1})}

\begin{document}

\preprint{APS/123-QED}

\title{Limits of Lateral Expansion in Two-Dimensional Materials with Line Defects}

\author{Pekka Koskinen}
 \email{pekka.j.koskinen@jyu.fi}
\affiliation{Nanoscience Center, Department of Physics, University of Jyv\"askyl\"a.}

\date{\today}

\begin{abstract}
The flexibility of two-dimensional (2D) materials enables static and dynamic ripples that are known to cause lateral contraction, shrinking of the material boundary. However, the limits of 2D materials' \emph{lateral expansion} are unknown. Therefore, here we discuss the limits of intrinsic lateral expansion of 2D materials that are modified by compressive line defects. Using thin sheet elasticity theory and sequential multiscale modeling, we find that the lateral expansion is inevitably limited by the onset of rippling. The maximum lateral expansion $\chi_\text{max}\approx 2.1\cdot t^2\sigma_d$, governed by the elastic thickness $t$ and the defect density $\sigma_d$, remains typically well below one percent. In addition to providing insight to the limits of 2D materials' mechanical limits and applications, the results highlight the potential of line defects in strain engineering, since for graphene they suggest giant pseudomagnetic fields that can exceed $1000$~T. 
\end{abstract}

\maketitle
The discoveries of two-dimensional (2D) materials were followed by reports of their subtle mechanical properties~ \cite{Akinwande2017}. They are never fully flat, since their elastic thinness make them susceptible for stabilizing out-of-plane rippling~ \cite{meyer_nature_07,lui_nature_09,Deng2016a}. Rippling also implies in-plane softening and considerable out-of-plane stiffening~ \cite{Blees2015,Kahara2020,Hiltunen2021}. However, materials' high in-plane stiffness keeps their surface area unchanged, which implies lateral contraction, shrinking of the material boundary~\cite{Nicholl2017}. This effect is best known from the negative thermal expansion coefficient of graphene~\cite{pozzo_PRL_11,balandin_nmat_11,Yoon2011}.


Rippling and lateral contraction are relevant for several reasons. Ripples affect substrate adhesion (and vice versa) as well as in-plane and out-of-plane deformations~\cite{Paronyan2011,Reddy2011,tapaszto_nphys_12,Koskinen2014,Lambin2014,Koskinen2018b}. They can be created by point defects~\cite{Zhang2014g}, adsorbates~\cite{thompson-flagg_EPL_09}, grain boundaries~\cite{malola_PRB_10,Lu2013}, or line defects~\cite{Kahara2020}, also without excessive hampering of material's mechanical and electronic properties~\cite{Zandiatashbar2014}. Contraction influences the functioning of nanoscale devices such as resonators and facilitates strain engineering to control both mechanical and electronic properties~\cite{Koskinen2014a,Deng2016a}. However, despite the prominence of rippling and contraction in practical applications and the abundance of related literature, one fundamental question remains open: \emph{what are the limits of intrinsic lateral expansion for 2D materials?}

An attractive strategy to address this question is to consider 2D materials with compressive line defects. The line defects can act as tiny stitches that can induce local stretched areas that---so the argument goes---cumulate into global lateral expansion~\cite{Kahara2020}. Representing various physical origins such as dislocations~\cite{Warner2013}, adsorbate arrays~\cite{thompson-flagg_EPL_09,Brito2011,Wang2017a}, stacking variations~\cite{Duerloo2014}, heterostructure interfaces~\cite{Wang2019a}, or grain boundaries~\cite{Ryder2016,Liu2010a}, line defects allow creating local compressive stress at relatively low defect density. While some line defects are created during material synthesis, others can be created afterwards by chemical means or even by direct laser irradiation~\cite{Koivistoinen2016,Johansson2017}. 

In this Letter I use thin sheet elasticity theory and sequential multiscale modeling to investigate the lateral expansion limits of 2D materials with compressive line defects. Theory permits analytical models with simple expressions for ripple properties and lateral expansion. It turns out that lateral expansion and rippling cannot coexist; rippling destroys expansion effectively.

\begin{table}[b]
    \centering
    \caption{Elastic parameters for selected 2D materials.}
    \begin{tabular}{lcccc}
         Material & $k_s$ (eV/\AA$^2$) & $k_b$ (eV) & $\nu$ & $t$ (\AA) \\
         \hline
         Graphene~\cite{kudin_PRB_01}         & 21  & 1.5 & 0.15 & 0.93 \\
         Bilayer graphene\cite{koskinen_PRB_10b} & 42  & 180 & 0.15 & 7.2 \\
         MoS$_2$~\cite{Cooper2013,Lorenz2012}          & 8   & 12  & 0.3  & 4.2 \\
         BN~\cite{kudin_PRB_01}               & 17  & 1.3 & 0.2  & 0.96 \\
         Silicene~\cite{C3RA41347K,Zhao2012}        & 3.8 & 0.4 & 0.4  & 1.1
    \end{tabular}
    \label{tab:material-table}
\end{table}

To model the defected 2D materials, I invoke the thin sheet elasticity theory~\cite{landau_lifshitz}, because it has proven effective and reliable even for atomic-scale deformations~\cite{kudin_PRB_01,Bao2009,Shenoy2010,koskinen_PRB_10,kit_PRB_12,Korhonen2014a,Memarian2015,Koskinen2016,Akinwande2017}. Theory characterizes membranes by bending modulus $k_b$, Poisson ratio $\nu$, and 2D Young's modulus $k_s$. Membrane's intrinsic length scale is given by the elastic thickness
\begin{equation}
    t=\sqrt{12 k_b/M},
    \label{eq:t}
\end{equation}
where $M=k_s/(1-\nu^2)$ is the longitudinal modulus. Elastic thickness equals the physical thickness of the 2D material when viewed as a slab of isotropic elastic membrane. Table~\ref{tab:material-table} shows parameters for selected 2D materials~\cite{kudin_PRB_01,koskinen_PRB_10b,Cooper2013,Lorenz2012,C3RA41347K,Zhao2012}.

\begin{figure}[t]
    \centering
    \includegraphics[width=\columnwidth]{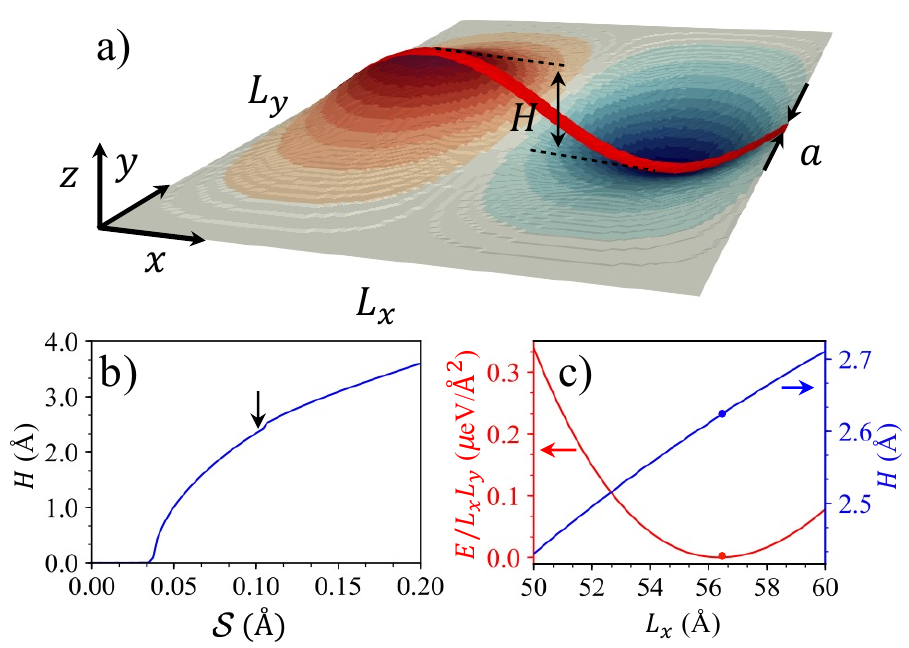}
    \caption{Rippling of $5$~nm$\times 10$~nm 2D membrane ($t=1.0$~\AA) by an infinitely long compressive line defect. a) One ripple wavelength from a line defect with $\mathcal{S}=0.1$~\AA\ ($a=2.5$~\AA\ and $\varepsilon_0=4$~\%). Vertical dimension is scaled by a factor of five and the width of the line defect is exaggerated. b) Ripple height $H$ as a function of the line defect strength $\mathcal{S}$. The membrane buckles at $\mathcal{S}=0.04$~\AA; the arrow points the geometry in panel a. c) Surface energy density (left scale) and ripple height (right scale) as a function of ripple wavelength $\lambda=L_x$ for $\mathcal{S}=0.1$~\AA.}
    \label{fig:1}
\end{figure}

The theory can be augmented to include compressive line defects, modeling them as stripes of width $a$, length $l$, and a pre-strain $\varepsilon_0$ that implies the equilibrium length $l(1+\varepsilon_0)$ (Fig.~\ref{fig:1}a)~\cite{Kahara2020}. The magnitude of $a$ is around $2-3$~\AA\ as it arises from the atomic structure of the defect~\cite{Kahara2020}. The parameter
\begin{equation}
\mathcal{S}=a\varepsilon_0 
\label{eq:S}
\end{equation} 
characterizes the \emph{strength} of the line defect. For small deformations (strains $\ll \varepsilon_0$) the line defect corresponds to one-dimensional line stress $\tau=M\mathcal{S}$. The strength $\mathcal{S}$ is unique for given 2D material and line defect, but here I treat it as a continuous parameter. I ignore tensile pre-strain ($\varepsilon_0<0$), because it cannot induce lateral expansion in any situation. Based on earlier atomic simulations, reasonable compressive pre-strains lie in the range $\varepsilon_0 \lesssim 20$~\%~\cite{Kahara2020}.

The theory was then harnessed for numerical simulations of defected membranes in an $L_x \times L_y$ periodic rectangular cell. Membrane was discretized to an $N_x \times N_y$ grid and the optimum morphology was solved numerically by minimizing the total elastic energy; see Supplemental Material (SM) for details~\cite{SM}. Materials of different elastic thicknesses $t=1.0\ldots 10$~\AA\ were simulated by adopting a fixed Poisson ratio ($\nu=0.15$) and longitudinal modulus ($M=21.5$~eV/\AA$^2$) while varying $k_b$ according to \eq{eq:t}. Line defect strengths $\mathcal{S}$ were adjusted by choosing the width equal to a typical lattice constant $a=2.5$~\AA\ and varying $\varepsilon_0$. Since the main parameters are $t$ and $\mathcal{S}$, the above choices do not restrict the general validity of the results. In the numerical implementation, because the atomic scale is much smaller than the grid spacing ($a \ll L_x/N_x$), the line defects were introduced via a sequential multiscale model (SM)~\cite{SM}. 

To construct a comprehensive understanding of the effect of line defects, I start by discussing isolated infinite and finite line defects before analyzing experimentally relevant random line defect networks.


Consider a $5\text{ nm}\times 10\text{ nm}$ membrane with $t=1.0$~\AA\ and an infinitely long ($l=L_x$) line defect along $x$-axis (\fig{fig:1}a). The membrane is initially planar, but buckles to a rippled conformation upon increasing line defect strength from zero to $\mathcal{S}=0.04$~\AA\ (Figs~\ref{fig:1} a~and~b). The ripple forms because it releases the compressive stress of the line defect. Further increase in $\mathcal{S}$ leads to monotonous increase in ripple height. The ripple height profile can be approximated by the sine wave
\begin{equation}
z(x,y)=\tfrac{1}{2}H \exp{\left(-y^2/2\sigma^2\right)}\sin{(2\pi x/\lambda)},
    \label{eq:sine}
\end{equation}
where $H$ is the peak-to-peak height, $\lambda$ is the wavelength, and $\sigma$ is a measure for the lateral width of the ripple. 

The above choice of $L_y=10$~nm was irrelevant because the ripple decays exponentially in $y$-direction. However, the choice of $L_x$ must be investigated in detail, as it directly determines the ripple wavelength. Fixing $\mathcal{S}=0.1$~\AA\ and increasing $L_x$ leads to monotonously increasing ripple height and a minimum of the surface energy density $E/(L_xL_y)$ at $L_x=56.3$~\AA\ with $H=2.62$~\AA\ (\fig{fig:1}c). This minimum implies that the ripple wavelength in an extended system is $\lambda=56.3$~\AA.  

\begin{figure}[t]
    \centering
    \includegraphics[width=\columnwidth]{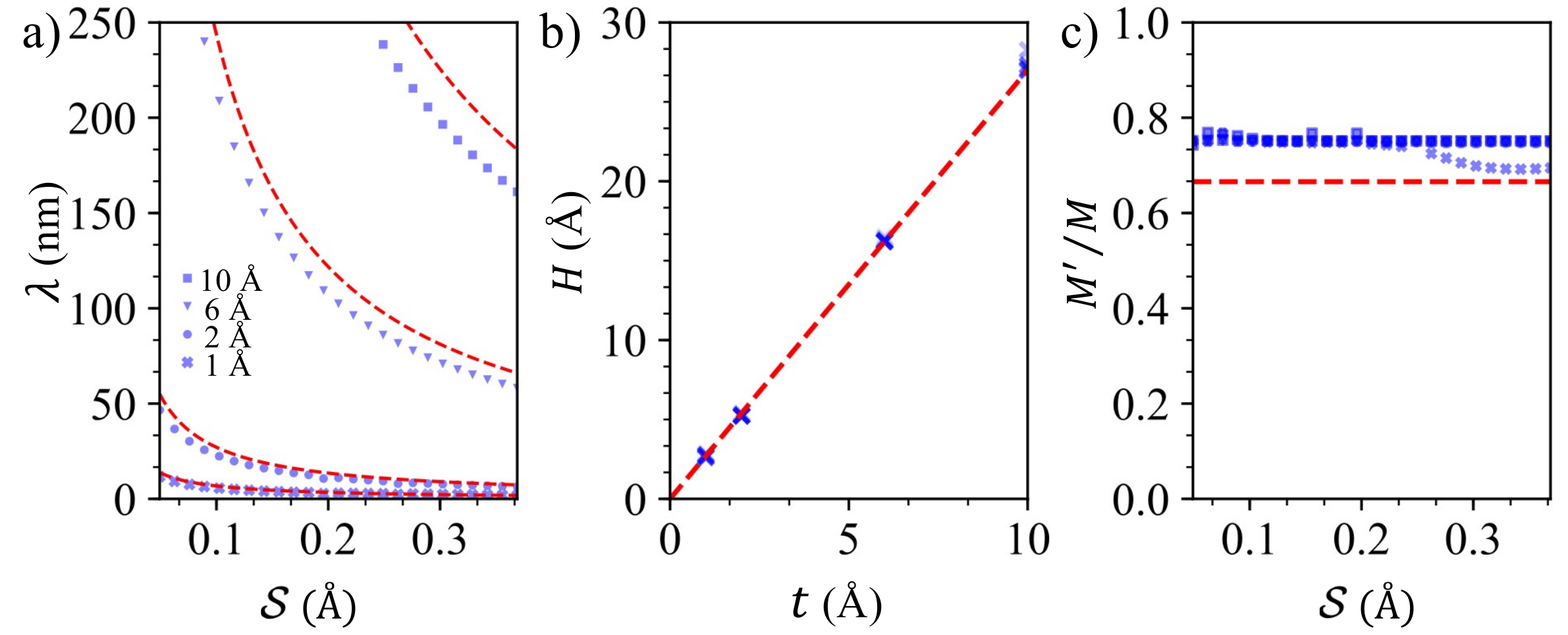}
    \caption{Ripple trends with infinite line defects. a) Optimum ripple wavelength as a function of line defect strength for different $t$. b) Ripple height as a function of elastic thickness. c) Longitudinal constant $M'$ of the entire simulation cell as a function of line defect strength. All panels show both numerical simulations (symbols) and analytical estimates [dashed lines from Eq.~(\ref{eq:lambda}) for $\lambda$, Eq.~(\ref{eq:H}) for $H$ and Eq.~(S14) for $M'$]. $L_x=\lambda$ and $L_y=1.5\cdot \lambda$ in all panels.}
    \label{fig:2}
\end{figure}

The rippling with line defects can be investigated also analytically. As derived in SM, adopting the ripple profile (\ref{eq:sine}) leads to the estimates for the optimum wavelength as
\begin{equation}
    \lambda=6.8\cdot t^2/\mathcal{S},
    \label{eq:lambda}
\end{equation}
for the ripple height as,  
\begin{equation}
    H=2.7\cdot t
    \label{eq:H}
\end{equation}
and for the ripple width as $\sigma\approx \lambda/5$~\cite{SM}. The estimates suggest that wavelengths increase for elastically thicker membranes and weaker line defects, which is plausible when viewed in terms of energy; shorter ripples require more energy, which is available in stronger line defects. Unexpectedly, however, the ripple height $H$ depends \emph{only} on elastic thickness and is independent of the properties of the line defect. This implies that ripples would form even with very weak line defects---although with very long wavelengths. In addition, the analytical model provides estimates for maximal slopes $\max(|dz/dx|)=1.2\cdot \mathcal{S}/t$, curvatures $\max(\mathcal{C}_{xx})=\mathcal{S}^2/t^3$, and strains $\max(\varepsilon_{xx})=0.4\cdot (\mathcal{S}/t)^2$ (SM)~\cite{SM}. For graphene, the strain field implies pseudomagnetic fields equal to $5\cdot 10^4\cdot \mathcal{S}^3$~\AA$^{-3}$T (SM)~\cite{SM,castro_neto_RMP_09,kim_EPL_08,Hsu2020}. For example, graphene with $\mathcal{S}=0.3$~\AA\ suggests maximal slopes $0.36$, curvatures $0.27$~nm$^{-1}$, local strains $3$~\%, and pseudomagnetic field that exceeds $1000$~T.




 
The analytical results are confirmed by systematic numerical simulations with $t=1...10$~\AA\ and $\mathcal{S}=0...0.37$~\AA. As Eqs.~(\ref{eq:lambda}) and (\ref{eq:H}) predict, ripple wavelengths are inversely proportional to $\mathcal{S}$ and quadratically proportional to $t$ (\fig{fig:2}a), while ripple amplitude is directly proportional to the elastic thickness, independent of $\mathcal{S}$ (\fig{fig:2}b).

This $\mathcal{S}$-independence of $H$ brings about a curious effect for lateral elastic properties. The longitudinal modulus $M'$ of the entire simulation cell, which accounts also for rippling, becomes entirely constant---independent of either $t$ or $\mathcal{S}$ (\fig{fig:2}c). Governing the energy curvature upon straining $L_x$, the longitudinal modulus depends on the width of the simulation cell, which here is $L_y=1.5\cdot L_x$. The constancy of $M'$ can be understood as follows: on one hand larger $t$ increases rippling height [\eq{eq:H}] and thereby tends to decrease $M'$, but on the other hand larger $t$ increases bending stiffness and thereby tends to increase $M'$. Combined, these two tendencies cancel and $M'$ becomes approximately constant. An analytical calculation gives the estimate $M'/M=\left( 1-0.5\cdot\lambda/ L_y\right)$ which becomes $0.67$ for current parameters (\fig{fig:2}c; SM)~\cite{SM}.

\begin{figure}
    \centering
    \includegraphics[width=\columnwidth]{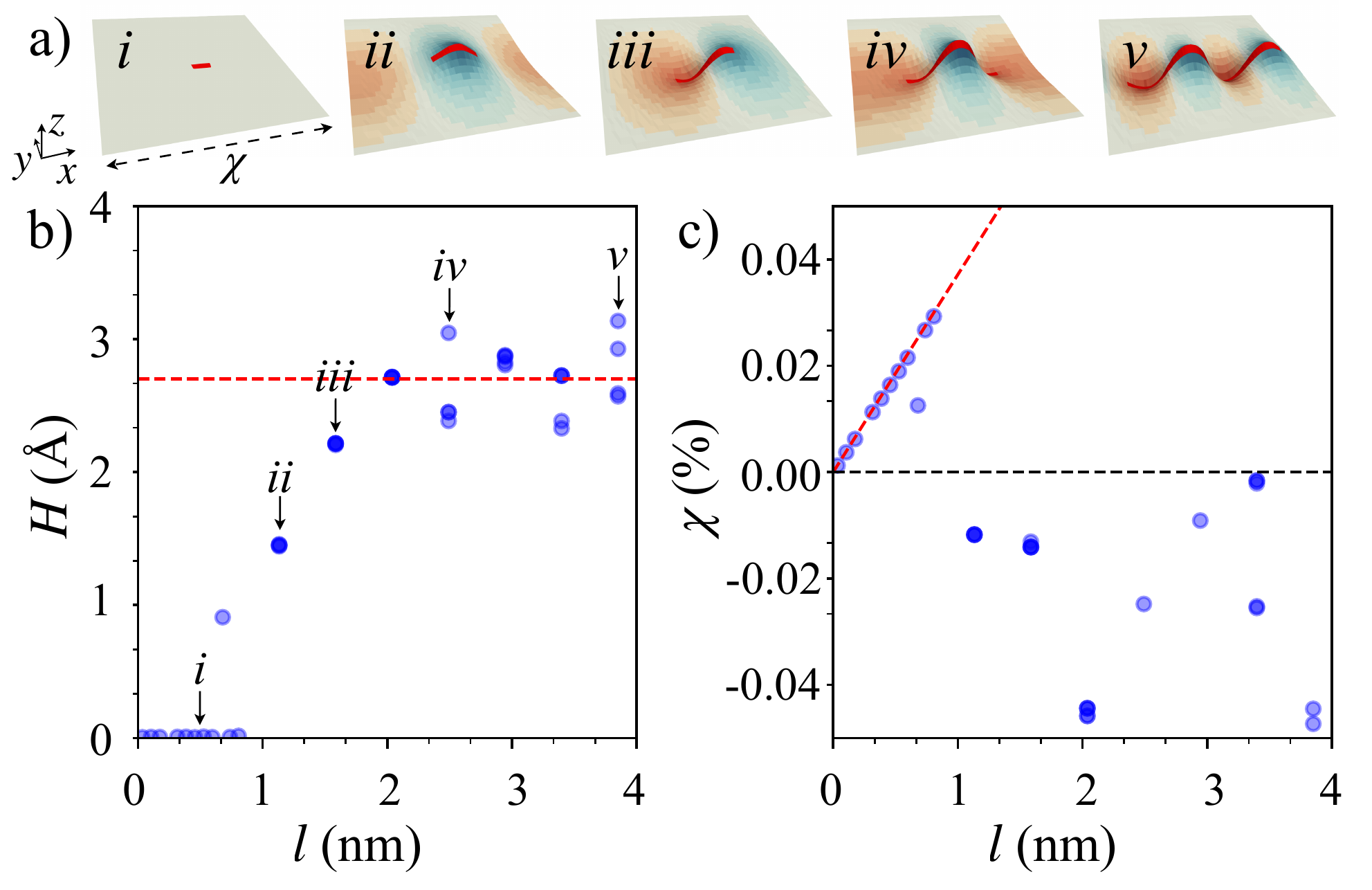}
    \caption{Lateral uniaxial expansion of $10$~nm$\times 6.6$~nm membrane with $t=1.0$~\AA\ for line defects with $\mathcal{S}=0.25$~\AA\ and finite length $l$. a) Development of ripple morphology with increasing $l$. The vertical dimension is scaled by a factor of five (see panel b for the color scale). b) Ripple height as a function of $l$. The arrows point at geometries in panel a. The red dashed line shows the ripple height estimate from \eq{eq:H} for infinitely long defects. c) Lateral expansion of the membrane in $x$-direction. The red dashed line is the estimate from \eq{eq:ex1}. Fluctuations for given $l$ are due to sampling of random initial guesses.}
    \label{fig:3}
\end{figure}

Infinitely long line defects with optimum-wavelength ripples imply that the stress in $x$-direction vanishes. Any residual stress would lead to strain that changes the wavelength, creating a contradiction with the presumption of an optimum wavelength. This notion implies an intermediate result: infinitely long (length$\gg\lambda$) line defects \emph{cannot} induce lateral expansion. But what about finite line defects?

To address this question, consider line defects whose lengths $l$ are around the optimal wavelength, $l\lesssim \lambda$. Let us fix $\mathcal{S}=0.25$~\AA\ with $t=1.0$~\AA\ and gradually increase the length $l$. Initially, at small $l$ the membrane remains flat, until at $l\approx 0.7$~\AA\ it buckles to form a single bump (\fig{fig:3}a). Simulations contain fluctuations due to random initial guesses. Analytical model similar to the one of infinite line defects gives the scaling $l_b=2.1\cdot t^2/\mathcal{S}\approx 0.8$~\AA\ for the buckling limit, in fair agreement with numerical simulations (SM)~\cite{SM}. After buckling, further lengthening leads to increased ripple height and development of alternating up-and-down bumps that gradually resemble the optimum ripple of the $l\gg \lambda$ limit (\fig{fig:3}a).

Yet, unlike infinite line defects, finite line defects \emph{can} induce lateral expansion. The expansion is defined as $\chi=(L_x-L_x^0)/L_x^0$, which is obtained by minimizing energy with respect to cell length $L_x$ for given initial length $L_x^0$. As the main observation, the membrane expands steadily upon increasing $l$ until it buckles (\fig{fig:3}c). The expansion is accurately described by the heuristic model
\begin{equation}
\chi=\mathcal{S}l/L_xL_y.
\label{eq:ex1}
\end{equation}
The model means that the hidden area of the line defect ($a\cdot l\varepsilon_0=l\mathcal{S}$) proportionally increases the surface area of the membrane ($L_xL_y$). With $l>l_b$ the membrane ripples and loses its capacity to sustain the compressive stress, rendering the expansion unpredictable.

\begin{figure}[tb]
    \centering
    \includegraphics[width=\columnwidth]{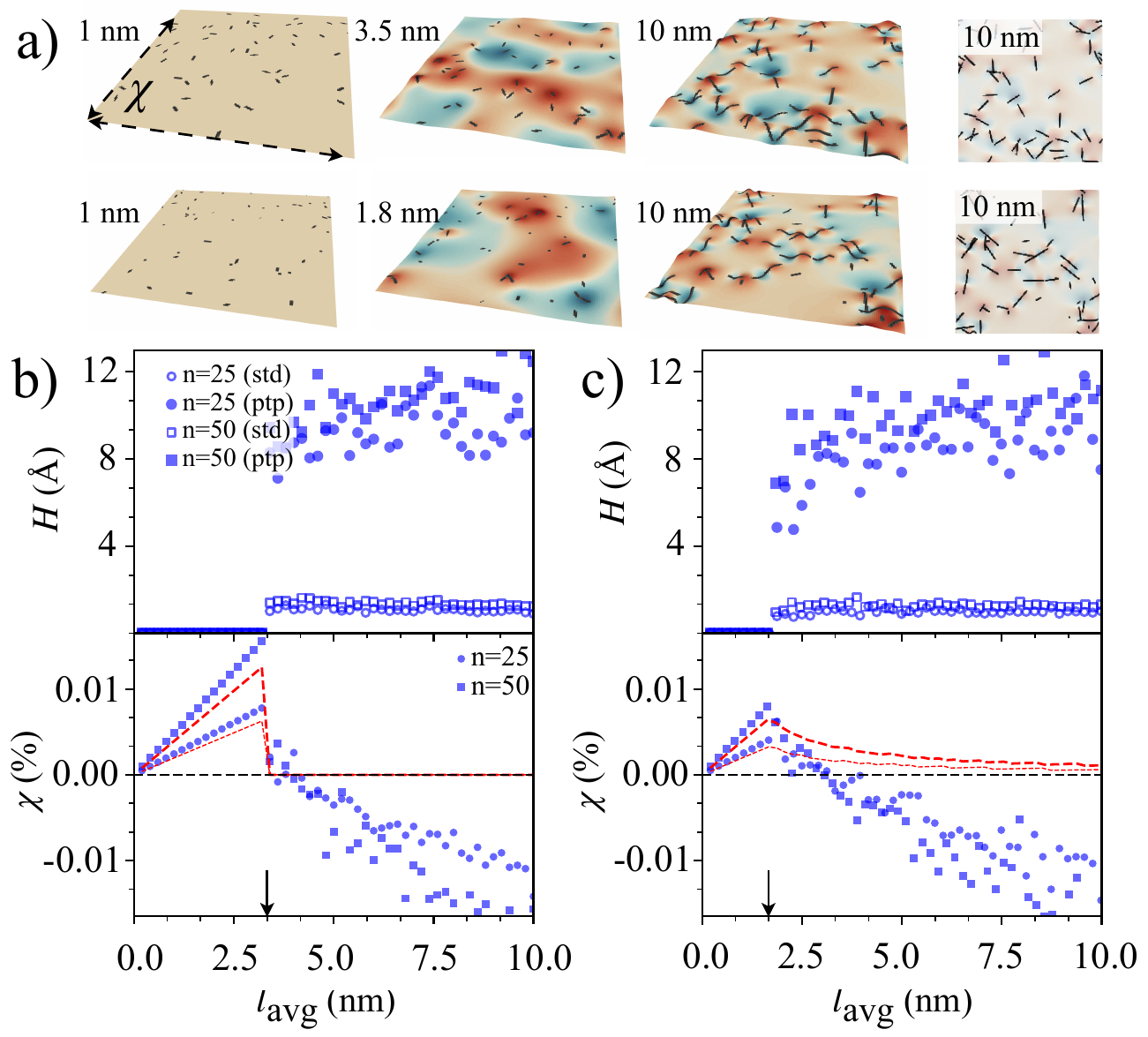}
    \caption{
    Lateral biaxial expansion of $100$~nm$\times 100$~nm membranes ($t=2.0$~\AA) filled by networks of line defects with $\mathcal{S}=0.25$~\AA. a) Ripple morphology snapshots with $50$ line defects. The networks have even (upper) and linear (lower) length distributions and increasing mean length (shown above). For clarity, the figures on the right show the networks with top views. b) Ripple peak-to-peak (ptp) and standard deviation (std) heights (upper panel) and biaxial strain (lower panel) as a function of mean defect length for $n=25$ and $n=50$ line defects with even length distribution. The red dashed line is the strain estimate from \eq{eq:ex}. c) Same as panel b for linear length distribution. Arrows point to mean lengths corresponding to the initial buckling of the longest line defects [Eq.~(S22)]; for even distribution $l_\text{max}=l_\text{avg}$ and for linear distribution $l_\text{max}=2l_\text{avg}$.}
    \label{fig:4}
\end{figure}

These results provide sufficient insight to proceed to realistic random line defect networks~\cite{Hiltunen2020}. I considered a $100$~nm$\times 100$~nm membrane with $n=25$ and $50$ randomly placed and oriented line defects of various lengths (\fig{fig:4}a). The corresponding densities ($0.25\cdot 10^{12}$~cm$^{-2}$ and $0.5\cdot 10^{12}$~cm$^{-2}$) are experimentally relevant and large enough for meaningful statistics but small enough to avoid excessive interaction between the line defects~\cite{Johansson2017}. The length distribution was either even ($l_i=l_\text{avg}$) or linear [$l_i=l_\text{avg}\cdot 2i/(n+1)$], where $i=1,2,\ldots n$ and $l_\text{avg}$ is the average length (\fig{fig:4}a). Such distributions can be justified by previous models~\cite{Hiltunen2020}.

For a defect network with even length distribution, the membrane expands laterally upon increasing $l_{avg}$ until the buckling threshold $l_\text{avg}>l_b$ (\fig{fig:4}b). After buckling the membrane ripples to a height that does not change much upon increasing $l_\text{avg}$ further. At the full $100$~nm scale the ripple height is $\sim 10$~\AA\, but at the local $\sim \lambda$ scale it is $\sim 5$~\AA, following \eq{eq:H}. For linear distribution the behavior is similar, only the transition to rippled membrane is less sudden. The gradual change occurs because individual line defects buckle at different $l_\text{avg}$. However, already the initial buckling of the longest defects ($l_\text{max}=2l_{avg}>l_b$) effectively eradicates the planar stress and destroys the lateral expansion. 

The lateral expansion is described accurately for both distributions by the generalization of \eq{eq:ex1},
\begin{equation}
\chi=\mathcal{S}l_\text{tot}(l<l_b)/L_xL_y,
\label{eq:ex}
\end{equation}
where $l_\text{tot}(l<l_b)$ is the cumulative length of all line defects below the buckling length $l_b$ (Figs~\ref{fig:4}b and \ref{fig:4}c). However, the membrane can sustain the lateral stress only as far as all defects remain below the buckling limit. After buckling the expansion becomes unpredictable. Ultimately, far beyond the buckling limit, the rippling strengthens and the membrane predominantly contracts~\cite{Liu2011}.

Thus, 2D materials can expand laterally only when line defects remain below the buckling limit of Eq.~(S22). The maximum expansion is reached when all defects have the maximum length $l_b$ and it equals 
\begin{equation}
\chi_\text{max}\approx 2.1\cdot t^2 \sigma_d,
\end{equation}
where $\sigma_d$ is the defect density. For instance, for $t=1$~\AA\ and $\sigma_d=10^{12}$~cm$^{-2}$ the maximum expansion is $0.021$~\%. A reasonable estimate for a optimal defect density can be obtained by assuming that one line defect occupies a minimum area of $\sim (2 l_b)^2$. This assumption yields the theoretical maximum for the strain as $
\chi_\text{max}\approx \mathcal{S}^2/8 t^2$. For graphene ($t=1.0$~\AA) and $\mathcal{S}=0.3$~\AA\ this implies $\chi_\text{max}\approx 1$~\%. 

Finally, I discuss briefly the role of substrates, which were excluded from the simulations. The transition from flat to rippled membranes reduces the energy by $0.13\cdot M\mathcal{S}^3/t^2$ per unit length of an infinite line defect (SM)~\cite{SM}. Assuming that the defects have an effective width of $\lambda/2$, this translates into surface energy density of $0.04\cdot M(\mathcal{S}/t)^4$. For $M\sim 20$~eV/\AA$^2$, $t=1.0$~\AA\ and $\mathcal{S}=0.35$~\AA\ the energy density becomes $\sim 10$~meV/\AA$^2$---and competes with a typical strength of van der Waals adhesion~\cite{koenig_nnano_11,Wang2016b,Duong2017}. Moreover, substrates themselves can be used for defect and strain engineering~\cite{10.1126/sciadv.aaw5593}. In short, a very strong adhesion can dominate membrane mechanics completely and effectively prevent both rippling and sliding. A very weak adhesion can allow for both rippling (desorption) and sliding, so that the rippling and expansion remains governed by membrane's intrinsic dynamics. However, an intermediate adhesion can suppress rippling but still allow sliding. For such an adhesion the rippling instability would not limit the maximal intrinsic expansion anymore; the expansion would still be given by Eq.~(\ref{eq:ex1}), but its upper limit would be given by maximum practical defect density.

To conclude, the onset of rippling dictates the limits of the lateral expansion of 2D materials. The theoretical maximum for the lateral expansion of the thinnest ($t=1$~\AA) materials is around $1$~\%, local strains being far greater. To diminish the effect of substrates, the expansion would be best measured experimentally from suspended 2D material samples or from customized 3D blisters of 2D materials, such as demonstrated for graphene by optical forging~\cite{Hiltunen2021}. The simulations and analytical models presented here provide a comprehensive picture of the mechanical behavior of 2D materials with line defects and reveal new theoretical limits to open new avenues and further advance the design and strain engineering of 2D materials.

\nocite{Harris2020,Virtanen2020,bitzek_PRL_06}



%

\end{document}